\documentstyle[preprint,aps,epsfig]{revtex}

\newcommand{\lsim}{\raisebox{0.3mm}{\em $\, <$} 
\hspace{-3.3mm} \raisebox{-1.8mm}{\em $\sim \,$}}
\newcommand{\gsim}{\raisebox{0.3mm}{\em $\, >$} 
\hspace{-3.3mm} \raisebox{-1.8mm}{\em $\sim \,$}}

\begin{document}

\draft

\preprint{}

\title{Can Long-Baseline Neutrino Oscillation Experiments Tell Us 
about Neutrino Dark Matter?}

\author{Hisakazu Minakata and Yoriaki Shimada}
\address{Department of Physics, Tokyo Metropolitan University \\
Minami-Osawa, Hachioji, Tokyo 192-03, Japan}

\date{January 1997}

\preprint{
\parbox{5cm}{
TMUP-HEL-9702\\
hep-ph/9701362\\
}}

\maketitle

%%%%%%%%%%%%%%%%%%%%%%%% ABSTRACT %%%%%%%%%%%%%%%%%%%%%%%%%%%%%
\begin{abstract}

We define the ratio $R$ = (No. of observed electron events)/
(No. of expected muon events $-$ No. of observed muon events)
as a measure for relative importance of
$\nu_\mu \rightarrow \nu_e$ and $\nu_\mu \rightarrow \nu_\tau$
oscillations which will be measurable in coming long-baseline
neutrino experiments. We then argue that if neutrino oscillation
is observed with 0.02 $< R <$ 0.87
it implies the rejection of the neutrino dark matter hypothesis.
Our argument is valid if neutrinos come with only three
flavors without steriles and if they have a hierarchical
mass pattern.

\end{abstract}

%%%%%%%%%%%%%%%%%%%%%%%%%%%%%%%%%%%%%%%%%%%%%%%%%%%%%%%%%%%%%%%

The possibility of neutrino dark matter is actively investigated 
both from theoretical and experimental points of view. 
The cold and hot dark matter cosmology is one of the rival models 
which can account for structure formation in the universe 
\cite {mixdm}. The hot dark matter is the indispensable ingredient 
in the scenario by which the normalization of the density fluctuation 
required by structure formation can be made consistent with the 
COBE observation \cite {WS}. 
The neutrinos with the mass range of 2 $-$ 20 eV are the best known 
candidate for the hot dark matter. 

The neutrinos with dark matter mass scale are subject to the active 
experimental searches. The ongoing short-baseline experiments, 
CHORUS \cite {CHORUS} and NOMAD \cite {NOMAD}, look for evidence 
for the neutrino oscillation 
$\nu_\mu \rightarrow \nu_\tau$ with the sensitivity 
$\sin^2 2\theta \simeq 3 \times 10^{-4} (3 \times 10^{-3})$ 
for the mass range $m_{\nu_{\tau}} \gsim 100 (10)$ eV. Moreover, 
the new experiment COSMOS with greater sensitivity are planned and 
may be able to run at sometime early in the next century. 

In this paper we address a question which is rarely asked; 
can we learn about dark matter neutrinos from the long-baseline 
neutrino oscillation experiments? An immediate answer to this 
question is likely to be ``clearly no''. The principal objective of 
the long-baseline neutrino experiments is to confirm the large-angle 
mixing of either 
$\nu_\mu \rightarrow \nu_\tau$ or 
$\nu_\mu \rightarrow \nu_e$ 
which is strongly suggested by the atmospheric neutrino anomaly 
\cite {Kam,IMB,soudan}. Hence they are designed to probe the 
mass-squared difference of the order of 
$\Delta m^2 \simeq 10^{-2}$eV$^2$. 
These mass difference seems to be too tiny to probe the dark matter 
mass scale $\sim 10$eV, apart from the effects of large $\Delta m^2$ 
averaged over many period. 

Nevertheless, we are going to claim in this paper that the long-baseline 
neutrino oscillation experiments do probe the dark matter mass neutrinos. 

Our original motivation for this investigation started with the 
following question. Suppose that CHORUS and NOMAD or even future 
new experiments fail to observe the appearance events 
$\nu_\mu \rightarrow \nu_\tau$. Can we conclude from these results 
that the hypothesis of $\tau-$neutrino dark matter is excluded? 
The answer to this question is no for one obvious and one  
not-so-obvious reasons. 
The obvious one is that it can exist with smaller mixing angle which 
is outside of the sensitivities of these experiments. 
The non-obvious reason, which is more serious, is that in the 
three-flavor mixing framework there is a region of mixing parameters, 
a large-$s_{13}^2$ region,  
that cannot be probed by the particular appearance experiment
$\nu_\mu \rightarrow \nu_\tau$ but is allowed by all the accelerator 
and the reactor experiments. This fact is noticed and displayed in 
Fig. 12 of \cite {FLS}. 
The latter reasoning is more serious because we cannot conclude for 
sure that the dark matter $\tau$ neutrino does not exist no matter 
how the sensitivity of the experiment is improved. 

These considerations naturally lead us to the question if 
additional constraints on dark matter neutrinos could emerge 
from other experiments in an unexpected way. In this context, 
we examine the long-baseline neutrino oscillation experiments 
which might be able to tell something about the dark matter 
neutrinos. We show in this paper that if the neutrinos come only 
with three flavors, i.e., without steriles, the long-baseline 
experiments can exclude the dark matter neutrino hypothesis 
unless their masses are strongly degenerate. We emphasize that 
to make such statement meaningful it is essential for the 
long-baseline experiments to have capability of doing detection 
not only in the $\nu_\mu \rightarrow \nu_e$ appearance channel 
but also in the disappearance channel. 

Since our conclusion heavily depends upon the theoretical framework 
let us define precisely what is the mass pattern and the three-flavor 
mixing framework with which we work throughout this paper. 
We assume that there exist three flavor of neutrino states with 
the mass hierarchy either 
$m_3 \gg m_1 \approx m_2$ or $m_1 \approx m_2 \gg m_3$. 
In each case 
$|\Delta M^2| \equiv |\Delta m^2_{13}|\approx |\Delta m^2_{23}|$ and 
is of the order of 5-100 eV$^2$. There is only one remaining mass 
difference, $\Delta m^2 \equiv \Delta m^2_{12}$, 
which can be used as the solar or 
as the atmospheric neutrino mass scale. If we choose the former 
option we will not be able to detect any signals in the long-baseline 
experiments and hence no useful constraint will be obtained. 
For this reason we employ the alternative option 
$\Delta m^2 \simeq 10^{-2}$eV$^2$, and {\it assume} that the 
experiments do observe the neutrino oscillations.  
Thus, we cannot accommodate the solar neutrino solution to our 
restrictive framework of three-flavor mixing enriched by mass 
scales suggested by the dark matter and the atmospheric neutrinos. 
We will return this point at the end of this paper. 

For definiteness we use the standard form of the Cabibbo-Kobayashi-Maskawa 
mixing matrix 

\begin{eqnarray}
U=\left[
\matrix {c_{12}c_{13} & s_{12}c_{13} &  s_{13}e^{-i\delta}\nonumber\\
-s_{12}c_{23}-c_{12}s_{23}s_{13}e^{i\delta} &
c_{12}c_{23}-s_{12}s_{23}s_{13}e^{i\delta} & s_{23}c_{13}\nonumber\\
s_{12}s_{23}-c_{12}c_{23}s_{13}e^{i\delta} &
-c_{12}s_{23}-s_{12}c_{23}s_{13}e^{i\delta} & c_{23}c_{13}\nonumber\\}
\right],
\label{eqn:CKM}
\end{eqnarray}
for the neutrino mixing matrix $\nu_\alpha = U_{\alpha j}\nu_j \;(\alpha=
e,\mu,\tau \;\mbox{and}\; j=1,2,3)$. 

If one of the two independent $\Delta m^2$ is of the order of dark 
matter mass scale the mixing angles are subject to the stringent 
constraint by the data of accelerator and reactor experiments. 
The clear recognition of this point in a general three-flavor setting 
has started with the works in refs. \cite{mina,BBGK}, and the 
quantitatively accurate bound is worked out by Fogli, Lisi, and 
Scioscia \cite {FLS} in the explicit CKM parametrization 
as ours (\ref {eqn:CKM}). The resultant allowed parameter region 
can be plotted on $s_{13}^2-s_{23}^2$ plane. (See Fig. 2.) There 
exist three allowed regions 
\begin{enumerate}
\renewcommand{\labelenumi}{(\roman{enumi})}
\item small-$s_{13}^2$ and small-$s_{23}^2$
\item large-$s_{13}^2$ and arbitrary $s_{23}^2$
\item small-$s_{13}^2$ and large-$s_{23}^2$
\end{enumerate}
It is of key importance to recognize that the tightly constrained 
region by short-baseline experiments can naturally lead to sizable 
oscillations at long-baseline \cite {mina}. Moreover, each 
allowed region (i)-(iii) corresponds to the pure two-flavor 
oscillation channels. Namely, the parameter regions 
(i), (ii) and (iii) correspond to almost pure 
$\nu_\mu \rightarrow \nu_e$, 
$\nu_\mu \rightarrow \nu_\tau$, and 
$\nu_e \rightarrow \nu_\tau$ oscillation channels, respectively. 

Then, it is very simple (but is crucial) to observe that any 
indication of genuine three-flavor mixing signals the failure of 
the dark matter mass hypothesis defined above. This is the task 
that the long-baseline neutrino oscillation experiments can do. 

To quantify the point let us focus on the particular type of the 
long-baseline experiment that can measure the appearance channel 
$\nu_\mu \rightarrow \nu_e$ as well as the disappearance channel, 
i.e., $\nu_\mu$ beam attenuation. 
We note that the experiment KEK-PS$\rightarrow$Super-Kamiokande, the 
only experiment that is already funded and will run in early 1999, 
is of this type. 

How can one characterize quantitatively the genuine three-flavor 
nature of the oscillation? 
We make a concrete proposal here. We propose the quantity 
\begin{eqnarray}
R &=&
\frac {\displaystyle \mbox{No. of observed electron events}}
{\displaystyle 
\mbox{No. of expected muon events} - \mbox{No. of observed muon events}}
\nonumber\\
%&&\hskip 2cm 
&=&
\frac{\displaystyle\int dE_\nu f(E_\nu)\sigma_{cc}^{(\nu_e)}(E_\nu)
P(\nu_\mu\rightarrow\nu_e)}
{\displaystyle\int dE_\nu f(E_\nu)\sigma_{cc}^{(\nu_\mu)}(E_\nu)
[1-P(\nu_\mu\rightarrow\nu_\mu)]}
\label{eq:R}
\end{eqnarray}
as the appropriate measure for the genuine three-flavor character of the 
oscillation. 
In (\ref {eq:R}) $f(E_\nu)$ indicates the muon neutrino flux at 
neutrino energy $E_\nu$ at the detector, 
$\sigma_{cc}^{(\nu_x)}$ denotes the charged current cross 
section of $\nu_x$ on targets in water, and 
$P(\nu_\alpha\rightarrow\nu_\beta)$ is the oscillation 
probability of $\nu_\alpha\rightarrow\nu_\beta$. 
 
If the oscillation is in almost pure $\nu_\mu \rightarrow \nu_e$ 
channel the ratio $R$ is close to unity. 
Whereas if it is in the almost pure $\nu_\mu \rightarrow \nu_\tau$ 
channel the ratio $R$ is approximately zero. 
Since the accelerator and the reactor bounds are so restrictive that tightly 
constrain the mixing parameters into the pure two-flavor oscillation 
channel the experimental observation of $R$ not too close to either 
zero or unity implies the rejection of the dark matter hypothesis. 
Our remaining task is to quantify this statement, which will be the 
topics of our subsequent discussions. 

We use approximate expressions of neutrino oscillation probabilities 
in vacuum as 
\begin{eqnarray}
P(\nu_\mu \rightarrow \nu_e) &=& 
 \frac{1}{2}s_{23}^2\sin^2 2\theta_{13}\nonumber\\
&+& \sin 2\theta_{12}c_{13}^2 \{ \sin 2\theta_{12}(c_{23}^2-s_{23}^2s_{13}^2)
+\cos 2\theta_{12}\sin 2\theta_{23}s_{13}\cos\delta \}\sin^2
(\frac{\Delta m^2 L}{4E})\\
&-& 2J \sin^1 (\frac{\Delta m^2 L}{2E})\nonumber
\end{eqnarray}

\begin{eqnarray}
1-P(\nu_\mu \rightarrow \nu_\mu) &=&
2s_{23}^2c_{13}^2(1-s_{23}^2c_{13}^2)
\nonumber\\
&+& 4(s_{12}^2c_{23}^2 + c_{12}^2s_{23}^2s_{13}^2 + \frac{1}{2}
\sin 2\theta_{12}\sin 2\theta_{23}s_{13}\cos\delta)\\
&\times& (c_{12}^2c_{23}^2 + s_{12}^2s_{23}^2s_{13}^2
-\frac{1}{2}\sin 2\theta_{12}\sin 2\theta_{23}s_{13}\cos\delta)
\sin^2(\frac{\Delta m^2 L}{4E})\nonumber
\end{eqnarray}
where $\Delta m^2\equiv \Delta m_{12}^2$ and 
$J=c_{12}s_{12}c_{23}s_{23}c_{13}^2s_{13}\sin\delta$ 
is the leptonic analogue of the Jarlskog parameter \cite {jarlskog}. 
These formulae are obtained by taking average over the rapid 
oscillations due to the large arguments in the trigonometric functions, 
$\displaystyle\frac{\Delta M^2 L}{4E} = 
127 (\frac {\Delta M^2}{1 eV^2})(\frac {L}{100 km})
(\frac {E}{1 GeV})^{-1}$, where 
$\Delta M^2 \equiv \Delta m_{13}^2 \simeq \Delta m_{23}^2$. 

We focus on the particular experiment KEK$\rightarrow$Super-Kamiokande 
and use the muon neutrino flux calculated by Nishikawa et al. 
\cite{nishikawa}. For the cross sections of the charged-current 
weak interaction we use the value obtained by Nakahata et al. 
\cite{nakahata}. We assume that the detection efficiency 
of electrons and muons in Super-Kamiokande is 100\%. 
This is a reasonable assumption because of high lepton energies 
$\gsim$ 200 MeV. For consistency check we have computed the number 
of events without neutrino oscillation. The expected number of 
muon events is 460-470 (whose uncertainty reflects errors in 
reading the neutrino flux off the figure in \cite {nishikawa}) 
for the neutrino flux from $10^{20}$ protons, 
which roughly corresponds to 12 month's run of the experiment, 
in rough agreement with the number $\sim$ 500 quoted by Nishikawa 
et al \cite {nishikawa}. 

We here make a comment on the earth matter effect. 
It is, of course, necessary to take into account the matter effect 
because it affects the oscillation probabilities at the level of 
a few to several \% depending upon the mixing parameters. 
However, we argue that for our purpose vacuum oscillation 
formula gives a fairly good approximation. The region of parameters 
which requires great accuracy is the regions with $R\simeq 0$ and 
$R\simeq 1$. In the former region the $\nu_\mu\rightarrow\nu_\tau$ 
oscillation is dominant and it is known that the matter effect is 
negligibly small in this region. In the latter region with 
$R\simeq 1$, on the other hand, the ratio $R$ is also insensitive 
to the matter effect. This is because it affects almost equally 
to the numerator and the denominator, since they are both dominated 
by the $\nu_\mu \rightarrow \nu_e$ oscillation. 
Therefore, ignoring the matter effect must give a fairly good 
approximation in computing $R$ in the relevant regions of mixing 
parameters for our purpose. 

In Fig. 1 we present bird's-eye view of the global features 
of the ratio $R$ and its projection onto the plane spanned by 
$\tan^2\theta_{13}$ and $\tan^2\theta_{23}$.   
The remaining angles are fixed as $\theta_{12}=\pi/4$ and 
$\delta =0$. $\Delta m^2$ is taken as $10^{-2} eV^2$. 
The correspondence between various symbols for lines 
and the equi-$R$ contours are exhibited in the figure. 
At small values of $\tan^2\theta_{13}$ and $\tan^2\theta_{23}$ 
$R \simeq 1$ as expected for 
$\nu_\mu\rightarrow \nu_e$ dominant region. 
(Unity of the cross section ratio 
$\sigma_{cc}^{(\nu_e)}$/$\sigma_{cc}^{(\nu_\mu)}$ 
at their effective region holds to the accuracy better than 
0.1 \%.) At large $\tan^2\theta_{13}$ $R \simeq 0$ as it should 
be for $\nu_\mu\rightarrow \nu_\tau$ dominant region.  

In Fig. 1(a) One notices that mountains are peaking out at 
$\theta_{23} \simeq \displaystyle\frac{\pi}{4}$ 
and $\theta_{13} \gsim \displaystyle\frac{\pi}{4}$. 
This feature can be understandable most easily at large 
$\tan^2\theta_{13}$ and $\theta_{12}=\theta_{13}= \pi/4$, where 
neutrinos can be made effectively unmixed after redefining 
the mass eigenstates. Therefore, both the numerator and the 
denominator of R are small and can have large ratios. If we vary 
$\theta_{12}$ away from $\pi/4$ the peak position moves to 
different values of $\tan^2\theta_{23}$ but the peak does not 
disappear. The feature may be regarded as a drawback of our 
definition of mixing angles in (\ref {eqn:CKM}) in discussing the 
$\nu_\mu\rightarrow \nu_\tau$ dominant region. 

Since we are interested in the region with sizable $\nu_\mu$ deficit 
we want to avoid to have a large $R$ value due to the artifact of the 
small denominator. Therefore, we impose the constraint 
$1-P(\nu_{\mu}\rightarrow\nu_{\mu}) \geq 0.1$ 
in our subsequent analysis. Namely, we set $R=0$ whenever the 
constraint is not met. We note that 10 \% deficit in probability 
amount to, very roughly speaking, more than 3 $\sigma$ away from 
the Kamiokande results \cite {Kam} which indicates about 40 \% 
deficit in the double ratio 
$(\frac {\nu_{\mu}}{\nu_{e}})_{observed}
/(\frac {\nu_{\mu}}{\nu_{e}})_{expected}$. 

The effect of the ``cut'' is displayed in Fig. 1(b); the mountains 
disappear. The cut also affects small-$\tan^2\theta_{13}$ and 
large-$\tan^2\theta_{23}$ region, the region of almost pure 
$\nu_{e}\rightarrow\nu_{\tau}$ oscillation. In fact, the ratio $R$ 
defined in (\ref{eq:R}) is not quite appropriate in discussing the 
$\nu_{e}\rightarrow\nu_{\tau}$ dominant region where $R$ is 
given by a ratio of two small numbers and hence its value is 
unstable. To explore this region of mixing angles we would need to 
repeat the similar consideration by using $\nu_{e}$ beam. In this 
paper we will confine ourselves to the $\nu_{\mu}$ experiment 
which is motivated by the atmospheric neutrino anomaly, leaving 
the discussion of $\nu_{e}$ experiment elsewhere. 

The ratio $R$ defined in (\ref {eq:R}) has a number of good features. 
It is independent of the absolute normalization of the neutrino flux 
which is difficult to determine in a good accuracy experimentally. 
It is very stable with respect to changes of the parameters such as 
$\Delta m^2$ and the CP-violating angle $\delta$.
We have checked that the Fig. 1 barely changes even if we take 
larger $\Delta m^2$, e.g., $\Delta m^2 = 5 eV^2$ 
instead of $\Delta m^2 = 10^{-2} eV^2$. The change is largest in 
regions with slopes, and smallest at the top and the bottom of the 
hill, the regions where are of interest to us. (See below.) But 
even the change in the former region is less than 0.1 \%, 
indicating the remarkable stability of the ratio $R$.   
We will later come back to the issue of stability against the 
change toward smaller value of $\Delta m^2$ and inclusion of 
the CP-violating angle. Until it will be done the value 
of $\Delta m^2$ is fixed as $10^{-2} eV^2$ throughout the 
computation in this work. 

In Fig. 2 we aim to present a global view of the equi-$R$ 
contours computed by varying the remaining angle $\theta_{12}$ from 
$10^\circ$ to $80^\circ$ in $1^\circ$ step at each point on the
$\tan^2\theta_{13}-\tan^2\theta_{23}$ plane. 
In the upper half plane ($\tan^2\theta_{13} > 1$) the 
contours of maximum values of $R$ obtained by varying 
$\theta_{12}$ are plotted for a given values of 
$\tan^2\theta_{13}$ and $\tan^2\theta_{23}$. 
In the lower half plane ($\tan^2\theta_{13} < 1$) the 
contours of minimum values of $R$ obtained by the same 
procedure are plotted. These prescriptions are to obtain the 
most conservative bound on $R$ in the respective regions of 
$\tan^2\theta_{13}$.  The contours are 
drawn in 0.1 units covering the region $R = 0.1 - 0.9$. 
The correspondence between various symbols for lines and the 
contours are again represented in the figure. 

One notices that there are two regions which receive an artificial 
erosion by the cut $1-P(\nu_{\mu}\rightarrow\nu_{\mu}) \geq 0.1$
imposed; one at large $\tan^2\theta_{13}$ and relatively small 
$\tan^2\theta_{23}$, the other at small $\tan^2\theta_{13}$ and 
large $\tan^2\theta_{23}$. The restriction to 
$10^\circ$ to $80^\circ$ for the angle $\theta_{12}$ is due to the 
fact that unless the restriction is imposed extra regions which 
do not meet the cut criterion are created, the situation outside 
of our physical interest. 

In Fig. 2 (and in Fig. 3) drawn by the thick lines are the 
constraints at the 90 \% CL imposed by the reactor and the accelerator 
experiments obtained by Fogli, Lisi and Scioscia \cite {FLS}.  
They are for $\Delta M^2 = 5 eV^2$. A comment is in order on the 
choice of this value. We want to derive the most conservative  
bound on the ratio $R$. Since the above constraint is milder for 
smaller $\Delta M^2$ we have to estimate the lowest possible 
value of the mass of the dark matter neutrinos. We use 
$m_{\nu}= 23.4 \Omega_{\nu} h_{50}^2/N_{\nu}$ \cite {mixdm}, where 
$\Omega_{\nu}$ is the fraction of neutrinos of the critical density, 
$N_{\nu}$ the number of species of dark matter mass neutrinos and 
$h_{50}$ denotes the Hubble parameter measured in units of 50 km/s 
$\cdot$ Mpc. We take a typical value $\Omega_{\nu} = 0.2$, 
the ``lowest'' value $h_{50} = 1$, and $N_{\nu} = 2$ to minimize 
$m_{\nu}$ within our ansatz of hierarchical mass pattern. We obtain 
$m_{\nu} = 2.34 eV$, hence our choice of $\Delta M^2 = 5 eV^2$.  

For our purpose of making quantitative statements on the ratio $R$ 
we need to know the detailed structure of the contours at 
$R \lsim 0.1$ and at $R \gsim 0.9$. It is presented in Fig. 3 where 
the equi-$R$ contours are drawn in 0.01 steps. 
The result presented in Fig. 3 quantifies our rough statement that 
the ratio $R$ must lie either close to zero or to unity in the 
presence of dark matter neutrinos; it must exist in restricted 
regions either $R < 0.02$ or $R > 0.87$, if the neutrinos come with 
only three flavors and the dark matter neutrinos have hierarchical 
mass pattern. 

We examine here the stability of $R$ against inclusion of the 
CP-violating angle and variation of $\Delta m^2$. 
We have verified, by computing the cases $\delta = \pi/4$, $\pi/2$ 
and $\pi$, that no appreciable change exists in equi-$R$ contours 
in the allowed regions of the mixing parameter plane. 
We have also checked that there is no detectable difference in 
the contours of $R$ in the allowed mixing parameter regions 
up to $\Delta m^2 = 5 \times 10^{-3} eV^2$. 
They cover interesting regions of $\Delta m^2$ as a solution 
of the atmospheric neutrino problem.   

A remark on the possibility of using other oscillation channels 
than $\nu_\mu\rightarrow \nu_e$ is in order. If the 
appearance measurement $\nu_\mu\rightarrow \nu_\tau$ is feasible, 
for instance in MINOS \cite{MINOS} and ICARUS \cite{ICARUS} 
experiments, one can do the same 
job by defining the similar ratio $R_{\tau}$ as 
$R_{\tau}$ = (No. of observed $\tau$ events)/
(No. of expected muon events $-$ No. of observed muon events). 
One then expect $R_{\tau} \sim 0$ at $\nu_\mu\rightarrow \nu_e$ 
dominant region and $R_{\tau} \sim 1$ at 
$\nu_\mu\rightarrow \nu_\tau$ dominant region.

Finally we make a comment on the limitation of our framework, the 
three-flavor neutrino mixing without steriles. Since the two 
$\Delta m^2$ are used up as the dark matter and the atmospheric 
neutrino mass scales, it cannot accommodate the solar neutrino 
solution. However, there exist partial resolution of the problem. 
If we are in the $\nu_\mu \rightarrow \nu_e$ dominant region the 
scenario does imply the deficit in the solar neutrino flux in the 
manner of Acker and Pakvasa \cite {AP}. Namely, the various 
neutrino flux is reduced by factor of $\sim$ 2 by the averaging effect 
due to the large $\Delta m^2$. On the other hand, if we are in the 
$\nu_\mu \rightarrow \nu_\tau$ region we would need to introduce 
a sterile neutrino which weakly mixes with $\nu_e$ and utilize 
the small-angle MSW mechanism \cite {MSW}. 

It is difficult to make definitive statements on how our result 
would change when we include a sterile neutrino in a generic way.  
We have to look for the quantities which can quantify relative 
importance of 
$\nu_\mu \rightarrow \nu_e$ and $\nu_\mu \rightarrow \nu_\tau$
oscillations analogous to our $R$. But, the one quantity would not 
be enough. So the discussion on the utility of the 
long-baseline neutrino experiments as a probe of the dark matter 
neutrino is entirely open even in the minimal three-flavor + 
one-sterile case. Fortunately, some models indicate the feature 
that the three-flavor neutrino sector is perturbed only weakly 
by the inclusion of the sterile neutrinos \cite {smirnov}. 
If this is the case our consideration in this paper could still 
apply at least approximately to such models with extra sterile 
species.

We have demonstrated that the long-baseline neutrino oscillation 
experiments can probe into the dark matter neutrinos. They can 
put scheme-dependent (such as three-flavor mixing without steriles) 
constraint and are capable of rejecting broad parameter regions 
associated with the dark matter hypothesis. To quantify such test 
of the hypothesis we have introduced the ratio $R$ as a measure 
for relative importance of $\nu_\mu \rightarrow \nu_e$ and 
$\nu_\mu \rightarrow \nu_\tau$ oscillations.    

We hope that the consideration in this paper urge long-baseline 
experimentalists even more strongly to design the experiments 
which allow them accurate measurement 
not only in the appearance but also in the disappearance channels.    

We thank K. Nishikawa for informative correspondences on the 
long-baseline experiment KEK-PS$\rightarrow$Super-Kamiokande.  
One of us (H.M.) has been supported in part by Grant-in-Aid for 
Scientific Research of the Ministry of Education, Science and Culture 
under \#0560355, and is supported by Grant-in-Aid for Scientific 
Research on Priority Areas under \#08237214.

%%%%%%%%%%%%%%%%%%%%%%%% Bibliography %%%%%%%%%%%%%%%%%%%%%%%%%

\newpage

\vspace{1.5cm}
\centerline{\large FIGURE CAPTIONS}
\vspace{0.5cm}

\begin{description}

\item[Fig.1.]
\begin{minipage}[t]{13cm}
\baselineskip=20pt
The bird's-eye view of the global features of the ratio $R$ and its 
projection onto the plane spanned by $\tan^2\theta_{13}$ and 
$\tan^2\theta_{23}$. The remaining parameters are taken as 
$\theta_{12}=\pi/4$, $\delta =0$ and $\Delta m^2 = 10^{-2} eV^2$.  
The correspondence between various symbols and the equi-$R$ contours
are exhibited in the figure. 
The figures (a) and (b) are for cases without and with constraint 
$1-P(\nu_{\mu}\rightarrow\nu_{\mu}) \geq 0.1$. (See the text.)

\end{minipage}
\vspace{0.5cm}

\item[Fig.2.]
\begin{minipage}[t]{13cm}
\baselineskip=20pt

Equi-$R$ contours are plotted on 
$\tan^2\theta_{13}-\tan^2\theta_{23}$ plane.  
In the upper (lower) half plane, $\tan^2\theta_{13} > 1$ 
($\tan^2\theta_{13} < 1$), the contours of maximum (minimum) 
values of $R$ obtained by varying $\theta_{12}$ are plotted 
for a given values of 
$\tan^2\theta_{13}$ and $\tan^2\theta_{23}$. 
$\Delta m^2$ is taken as $10^{-2} eV^2$. 
The correspondence between various symbols and the equi-$R$ contours
are represented in the figure. The thick solid lines are the 
constraints from reactor and accelerator experiments obtained by 
Fogli et al. \cite {FLS}. 

\end{minipage}
\vspace{0.5cm}

\item[Fig.3.]
\begin{minipage}[t]{13cm}
\baselineskip=20pt
The same as in Fig. 2 but the figure aims to display the detailed 
feature of the equi-$R$ contours at around the boundaries of the 
relevant parameter regions allowed by the terrestrial experiments.  
The large-$s_{13}^2$ and arbitrary $s_{23}^2$, and the 
small-$s_{13}^2$ and small-$s_{23}^2$ regions are presented 
in Fig.3(a) and Fig.3(b), respectively. The thick solid lines 
are the constraints from reactor and accelerator experiments 
obtained by Fogli et al. \cite {FLS}.

\end{minipage}
\vspace{0.5cm}

\end{description}

\newpage
\pagestyle{empty}
\epsfig{file=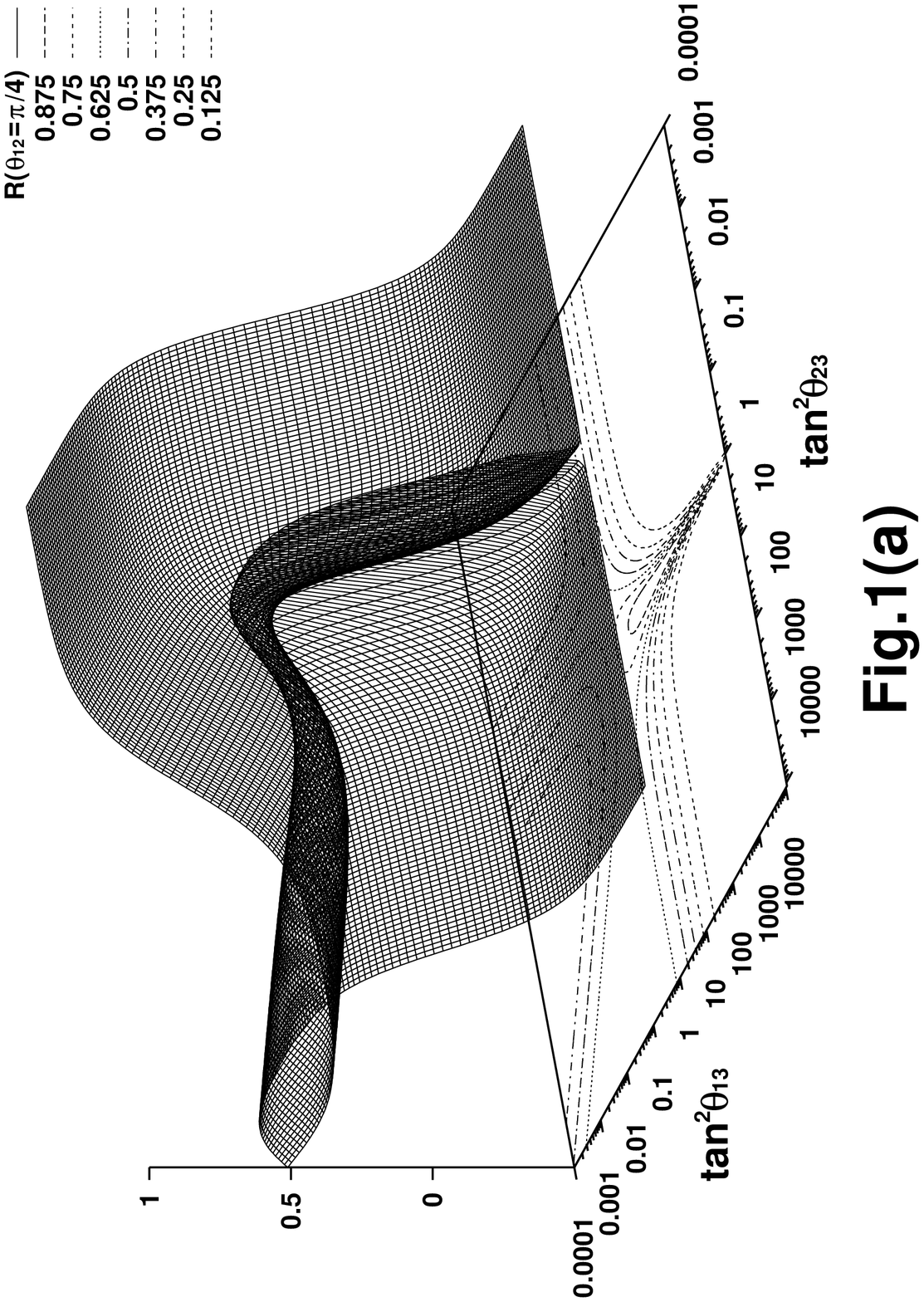}
\newpage
\epsfig{file=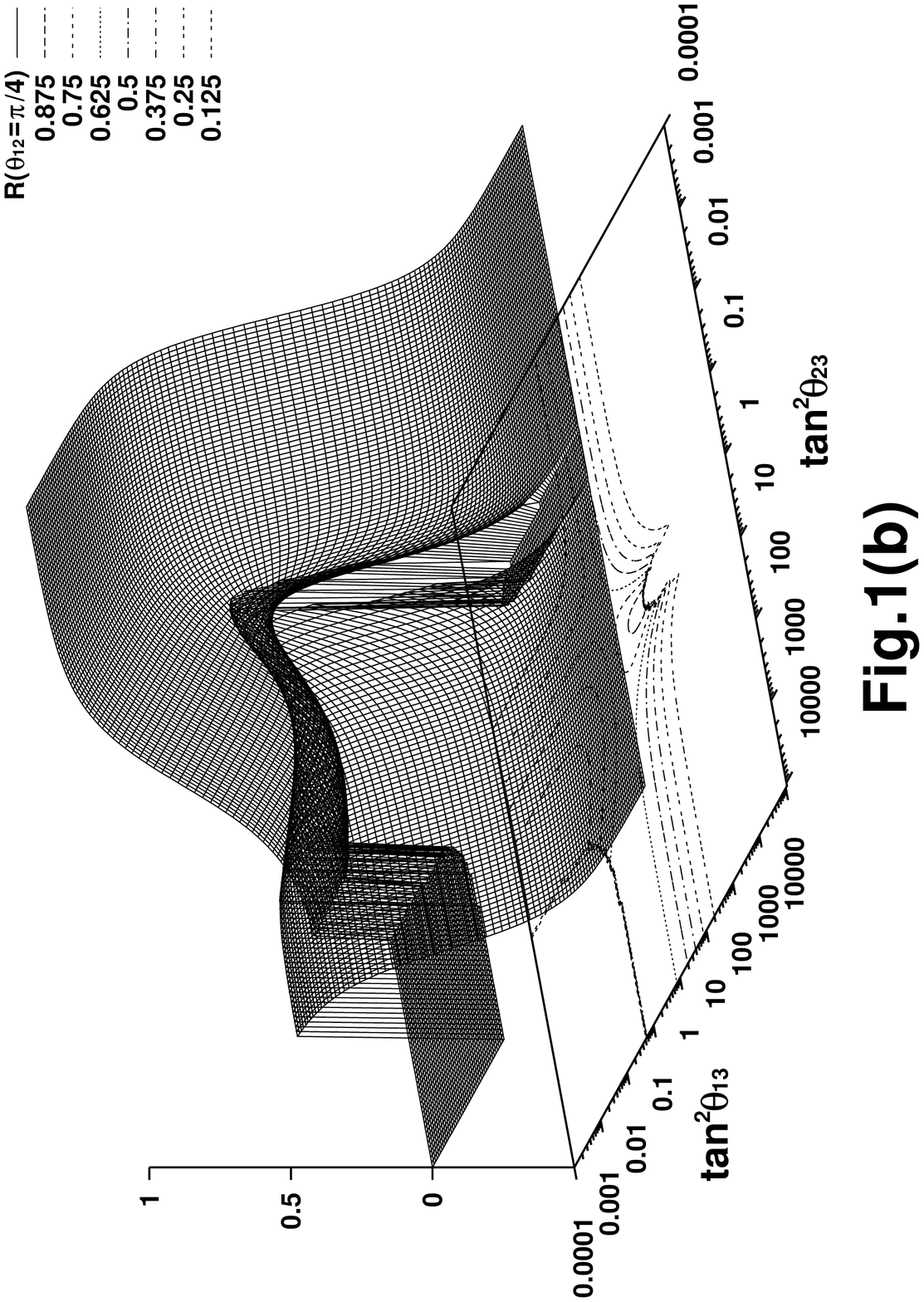}
\newpage
\epsfig{file=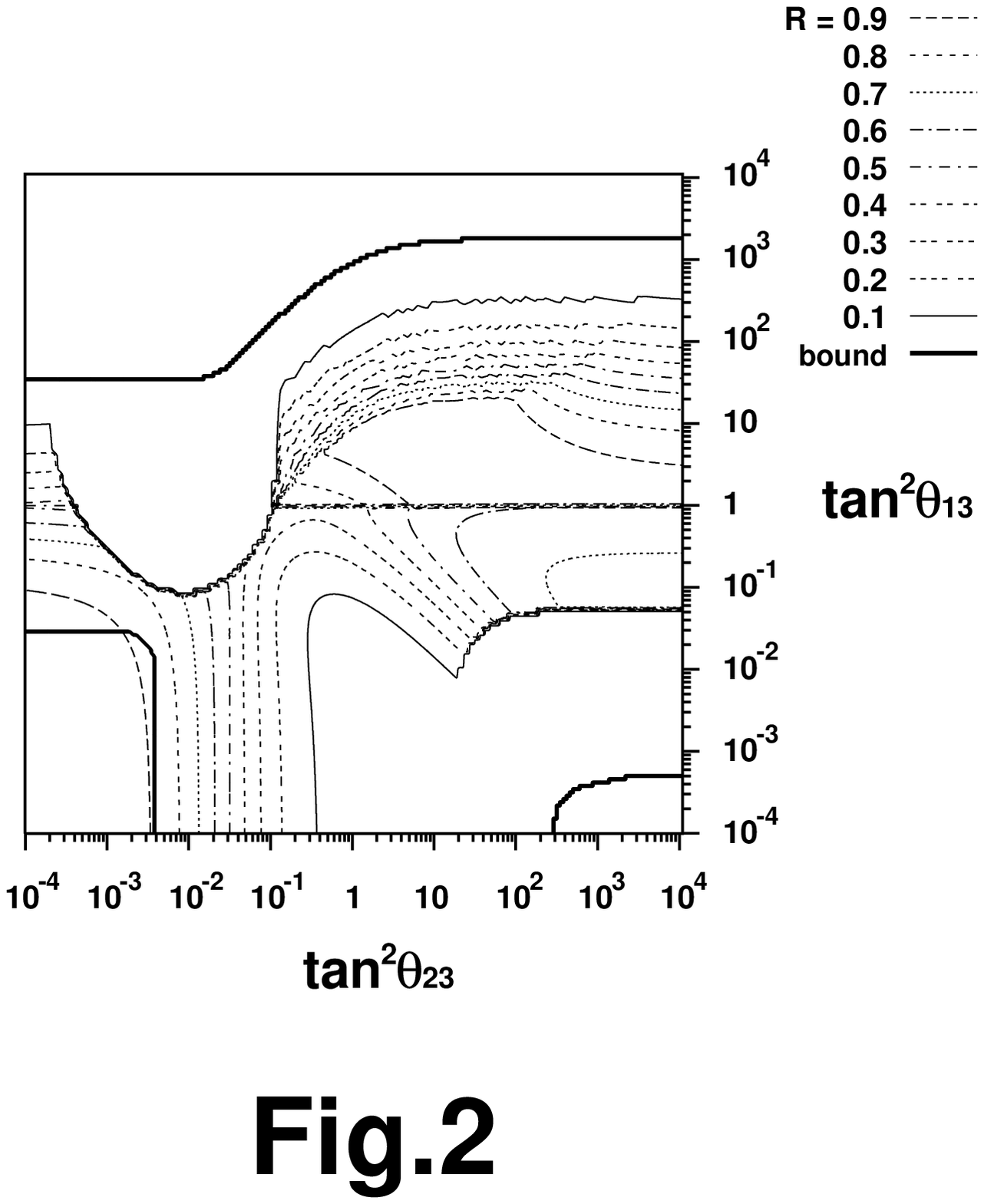}
\newpage
\epsfig{file=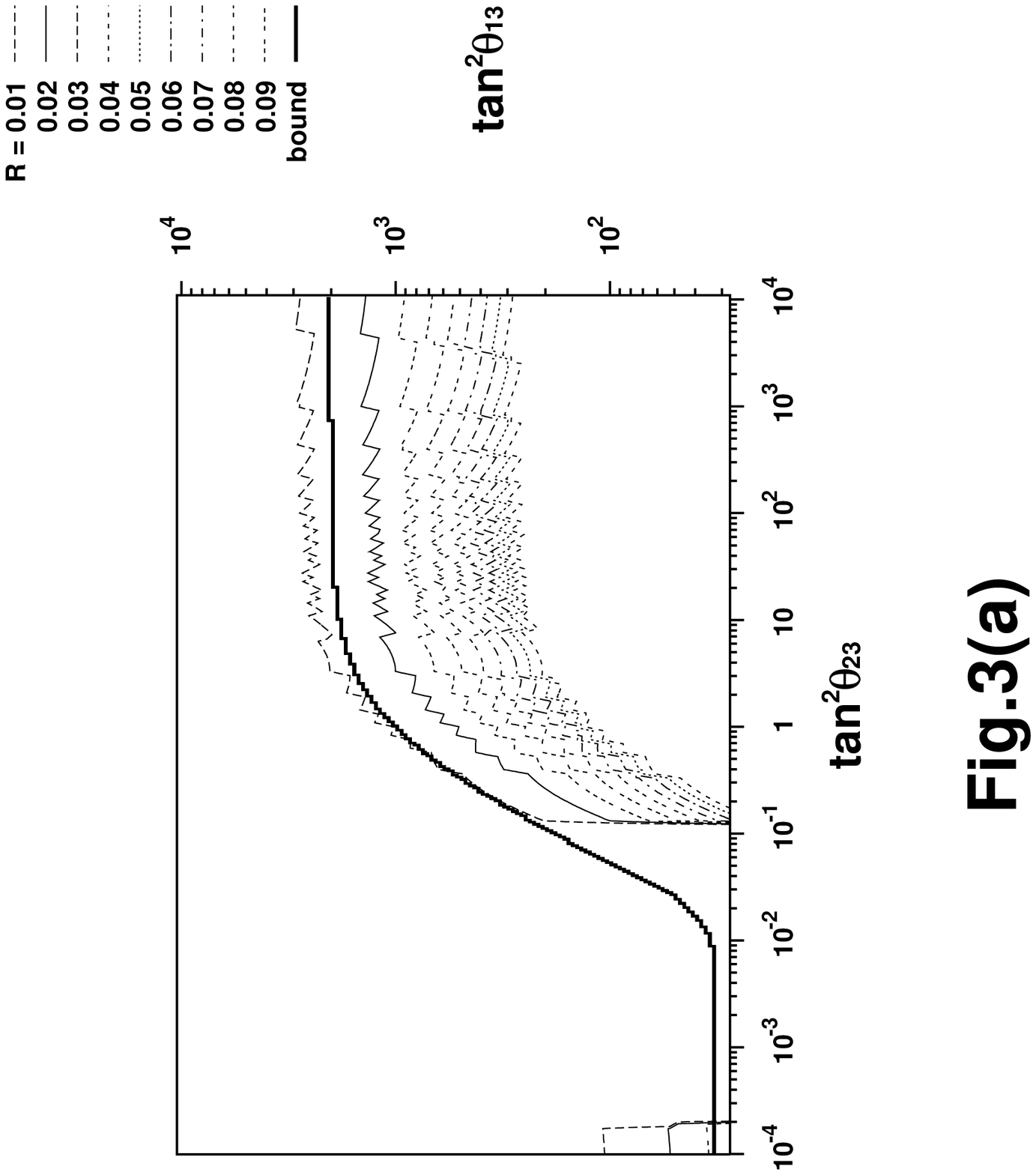}
\newpage
\epsfig{file=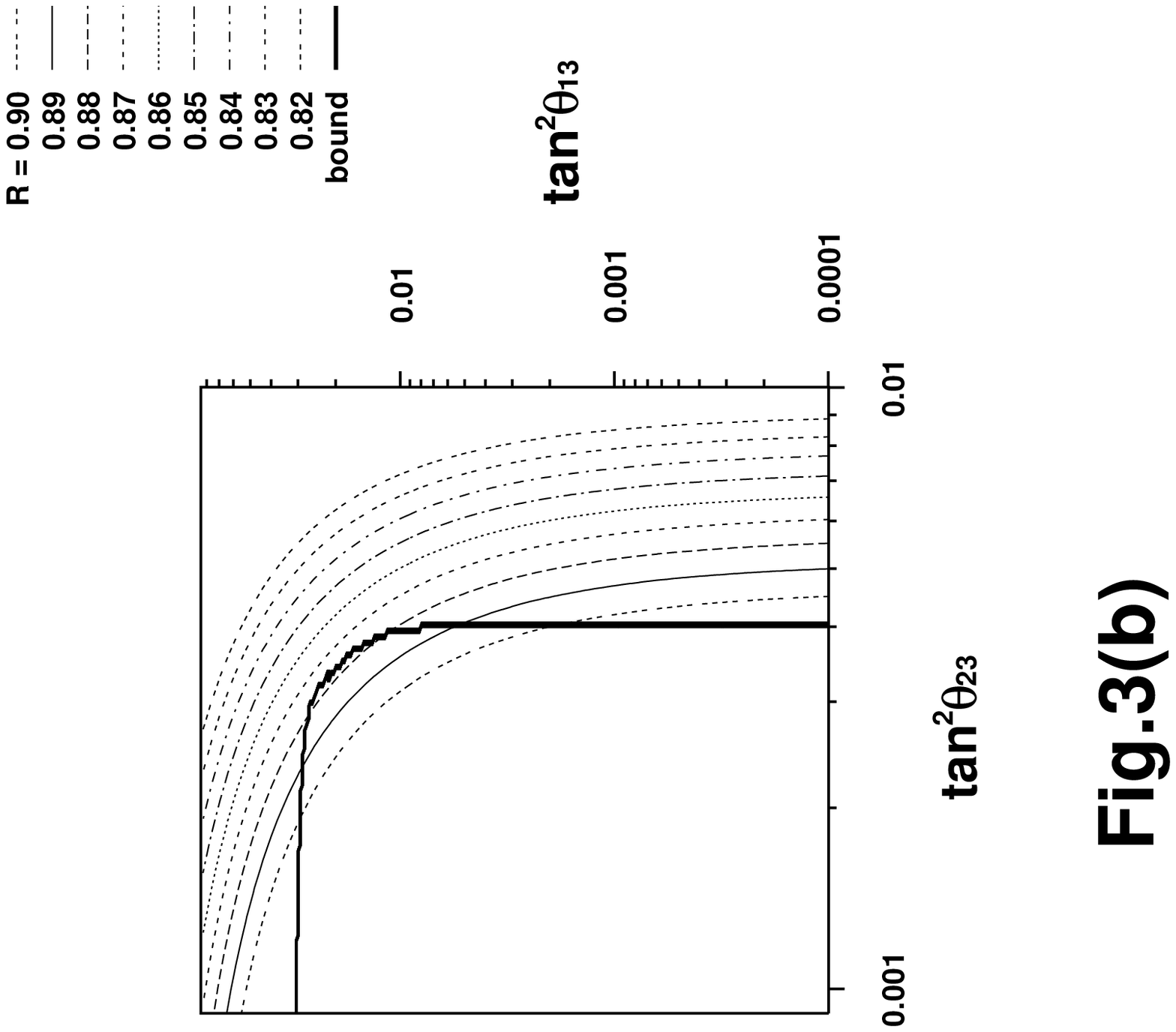}

\end{document}